\documentclass[runningheads]{llncs}

 \usepackage{eccv}

\usepackage{eccvabbrv}

\usepackage{graphicx}
\usepackage{booktabs}

\usepackage[accsupp]{axessibility}  % Improves PDF readability for those with disabilities.

\usepackage[pagebackref,breaklinks,colorlinks,citecolor=eccvblue]{hyperref}
\usepackage{orcidlink}

\begin{document}

\title{RMFA-Net: A Neural ISP for Real RAW to RGB Image Reconstruction} 

\titlerunning{Abbreviated paper title}

%\author{First Author\inst{1}\orcidlink{0000-1111-2222-3333} \and
%Second Author\inst{2,3}\orcidlink{1111-2222-3333-4444} \and
%Third Author\inst{3}\orcidlink{2222--3333-4444-5555}}
\author{Fei Li\inst{1}\orcidlink{0000-1111-2222-3333} \and
Wenbo Hou\inst{1}\orcidlink{1111-2222-3333-4444} \and
Peng Jia\inst{1}\orcidlink{2222--3333-4444-5555}}
\authorrunning{F.~Author et al.}

\institute{LiAuto, China \\
\email{\{lifei3,houwenbo,jiapeng\}@lixiang.com}}

\maketitle

\begin{abstract}
  Deep learning-based ISP algorithms have demonstrated significant potential in raw2rgb reconstruction. However, existing networks have not fully considered the specific characteristics of raw data, such as black level and CFA, which can negatively impact texture and color if mishandled. Moreover, uneven exposure in raw data is also not considered carefully, leading to adverse effects on contrast and brightness.
   In this paper, we introduce RMFA-Net to tackle these problems. We perform implicit black level correction to mitigate color shifts in dim scenes. To preserve high-frequency information and prevent misalignment, we propose a novel Three-Channel-Split mode. To address the issue of uneven exposure, we designed an explicit tone mapping module based on the Retinex theory.
   We train and evaluate our models using the dataset released by the Mobile AI 2022 Learned Smartphone ISP Challenge. It is demonstrated that RMFA-Net outperforms previous algorithms, achieving a PSNR score of over 25 dB, surpassing the state-of-the-art by +1 dB. Furthermore, we developed a lightweight version, RMFANet-tiny, for engineering deployment while still maintaining strong performance, surpassing the SOTA by +0.5 dB.
\end{abstract}

\section{Introduction}
\label{sec:intro}

The Image Signal Processor (ISP) is a specialized system designed to reconstruct RGB images from raw data captured by CMOS sensors. Traditional ISP algorithms are manually crafted and rely on a deep understanding of the sensors and complex tuning, which limits their applicability in fields such as autonomous driving and robotics. While image quality is well understood for human vision applications, it is not well-defined for visual perception systems. Deep learning-based ISP algorithms have emerged as a promising approach with significant potential and versatility. In recent years, there has been increasing interest in developing learning-based algorithms to design efficient and high-performance ISP algorithms tailored to specific domain requirements.
\begin{figure}[tb]
  \centering
  \includegraphics[width=1\linewidth]{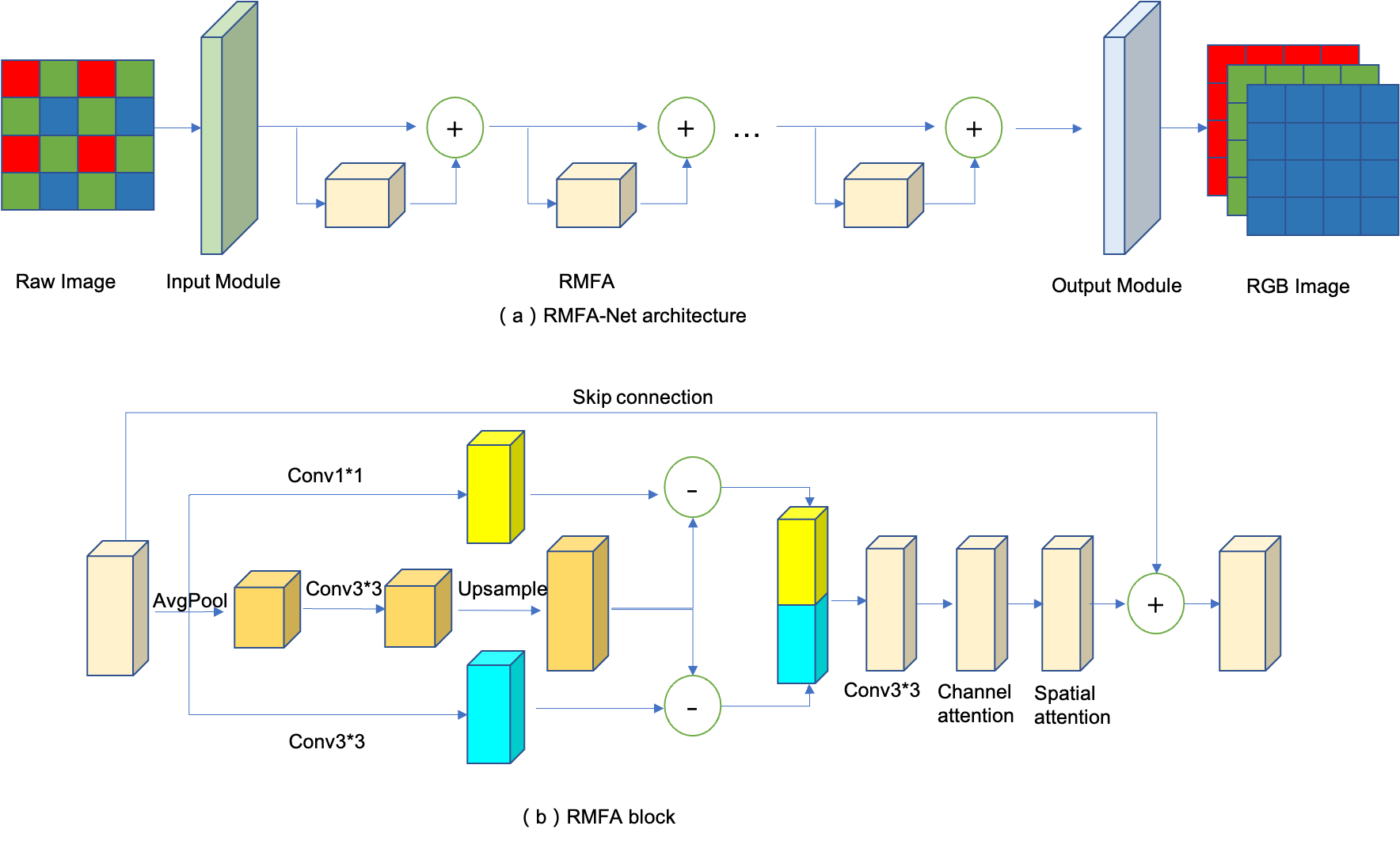}
  \caption{RMFA-Net: Residual-Multi-Frequency Attention Network. Figure (a) illustrates the overall architecture of RMFA-Net. The network consists of several key modules: the input module, stack of RMFA blocks and the output layer. Figure (b) illustrates  the internal structure of RMFA block.
  }
  \label{fig:networks}
\end{figure}
There are three primary approaches to enhance ISP performance with deep learning algorithms: image enhancement, network in the loop, and end-to-end neural ISP. Image enhancement algorithms~\cite{Dong2015,Cai2016,Guo2021,Kim2016,Kupyn2018} focus on improving specific aspects of image quality, such as denoising~\cite{Zhang2017}, HDR~\cite{Wang2021}, and super-resolution~\cite{Wang2020}, through post-ISP processing on RGB images. Network in the loop replaces certain sub-modules of traditional hardware ISP pipelines, like demosaicing~\cite{Liu2020}, with neural networks, necessitating frequent data exchange between different hardware types. Neural ISP is an emerging approach that replaces the entire ISP system with an end-to-end neural network~\cite{Schwartz2018}, directly taking raw data as input and producing RGB images, showing promise in enhancing image quality and perception performance.

When designing a neural ISP algorithm, careful consideration must be given to the preprocessing of raw data. We argue that raw data preprocessing should be sensor-dependent, considering characteristics such as black level, CFA, and pattern, which vary significantly among sensors~\cite{Lukac2005}. Black level, for instance, impacts color, particularly in dim scenes. Different CFAs have varying sampling rates. For example, in Bayer filter~\cite{Bayer1976}, the sampling rate of Green channel is twice of that of Red and Blue, which means the Green channel contains more information. Changes in patterns contribute to color shifts in trained networks. An effective ISP system should address key challenges related to resolution, contrast, color, and brightness. Deep learning perspectives suggest that resolution relies more on local features~\cite{Kong2022}, while color and brightness depend more on global features~\cite{Ignatov2020}. Global statistics affect overall image contrast, while local statistics impact contrast in smaller regions~\cite{Cai2018}. Therefore, a network should accurately extract and represent both global and local features.

In this work, we propose a novel data preprocessing method that preserves raw information by keeping the G channel unsplit while avoiding misalignment through pixel location invariant channel split. We introduce RMFA-Net (Residual Multi-Frequency Attention network), a new network architecture utilizing convolutional modules with various kernel sizes to extract global and local features. Additionally, we design an explicit tone mapping module based on Retinex theory~\cite{Land1971} to remove uneven exposure interference from raw data, facilitating more effective tone mapping learning. Finally, we combine these designs into a unified module called RMFA, serving as the fundamental building block of RMFA-Net. Through training and evaluation on the Fujifilm UltraISP dataset~\cite{Ignatov2023a}, our proposed method achieves image quality exceeding 25 dB in terms of PSNR and an SSIM score of 0.889.

Our contributions can be summarized as follows:

	[1]We propose a new method of data preprocessing that incorporates explicit black level correction to prevent color shifts in dim scenes. Additionally, we introduce a three-channel-split mode to address the differences in sampling rates, ensuring the preservation of complete raw information while avoiding misalignment; 
 
	[2]To improve the learning of tone mapping, we design an explicit algorithm based on Retinex theory, effectively removing uneven exposure from the raw data;
 
	[3]Furthermore, we present RMFA-Net, an end-to-end deep learning approach for raw to RGB image reconstruction. Through our proposed method, we achieve state-of-the-art image quality on the Fujifilm UltraISP dataset.

The remainder of the paper is structured as follows. In Section~\ref{sec:related_work} we summarized the related work on deep learning based isp algorithms from the angle of network architecture. Section~\ref{sec:proposed_method} presents our proposed network RMFA-Net and describes the underlying design choices. Section~\ref{sec:experiment} shows and analyzes the experimental results and discusses the limitations of the solution. Finally, Section~\ref{sec:conclusion} concludes the paper.

\section{Related Work}
\label{sec:related_work}
A seminal work in the field is PyNET~\cite{Ignatov2020}, an explicit multi-branch architecture model trained to directly map raw Bayer sensor data to RGB images captured by a DSLR camera. This model adopts an inverted pyramidal shape and processes images at five different scales, with each scale corresponding to a branch that is trained sequentially. PyNET effectively extracts both global and local features, merging them to generate the final outputs. It achieves image quality on par with commercial ISP system of Huawei P20 camera phone. Subsequent works, namely PyNET-CA~\cite{Kim2020}, introduced a channel attention mechanism to further enhance performance. Moreover, lightweight versions of PyNET, Micro ISP~\cite{Ignatov2023} and PyNET-V2~\cite{Ignatov2022}, were proposed to enable efficient execution on mobile devices, achieving a balance between image quality and computational efficiency.

Since the introduction of SRCNN~\cite{Kim2016}, which addressed the super-resolution challenge using CNN, various derivative models have been proposed to tackle image reconstruction tasks~\cite{Shi2016a, Zhang2018, Uchida2018, Chen2018}. Building upon this structure, AIISP~\cite{Ignatov2021} proposed the Channel Spacial Attention Network, incorporating double attention modules (DAM) with skip connections to enhance spatial dependencies and overcome the vanishing gradient problem. ENERZAi~\cite{Ignatov2021} introduced DenseNet-based residual blocks with separable convolutions, transpose convolution, and a channel attention mechanism to improve image quality. CVML~\cite{Ignatov2021} utilized residual blocks to extract a rich set of features from the input data, while the transposed convolution layer was employed for upsampling the final feature maps to the target resolution. EdS~\cite{Ignatov2021} proposed a ResNet-based architecture based on~\cite{Ignatov2017} and incorporated two additional $4\times4$ convolutional layers with a stride of 2 to extract global features. Multimedia~\cite{Ignatov2023a} introduced the enormous Re-parameter Convolution (eReopConv) layer as a replacement for standard convolution, while HITZST01~\cite{Ignatov2023a} proposed the Enhance Features Module in their RFD-CSA architecture, which effectively extracts features at multiple model levels while maintaining performance with a long-term residual connection

The U-Net~\cite{Ronneberger2015} architecture is also widely adopted in image reconstruction tasks~\cite{Wang2018, Isola2016, Cho2021, Jia2021, Hu2019}. Specifically for the Raw2RGB reconstruction task, the W-net~\cite{Uhm2019} combines two U-Net structures with channel attention modules and achieves good performance. SalGAN~\cite{Pan2017} employs the U-Net structure as the generator in an adversarial training scheme and incorporates a spatial attention scheme into the loss function. isp\_forever~\cite{Ignatov2021} proposes a U-Net-based model augmented with a channel attention module. MiAlgo~\cite{Ignatov2023a} introduces a 4-level UNet-based structure, where several convolutional layers are replaced with a residual group to enhance the network's reconstruction ability. CASIA 1st~\cite{Ignatov2023a} adopts a teacher-guided training strategy and proposes both teacher and student networks based on the U-Net architecture, incorporating a self-attention module

A notable event in the field is the Learned Smartphone ISP on Mobile GPUs with Deep Learning, Mobile AI \& AIM Challenge, which took place in 2021~\cite{Ignatov2021} and 2022~\cite{Ignatov2023a}. The challenge aimed to foster the development of efficient and high-performance models for inference on edge devices. Participants presented numerous innovative ideas that pushed the performance boundaries to new heights.

\section{Proposed Method}
\label{sec:proposed_method}
\subsection{Network Architecture}
\cref{fig:networks} provides a schematic representation of the proposed deep learning architecture. The network is divided into three main parts: the input module, a stack of RMFA blocks, and the output module. The input module takes images of size $256\times 256\times 3$ as input and expands the depth from 3 to a uniform width. In this part, two convolutional layers with a kernel size of $3\times 3$ are stacked. It is important to note that the tanh function is used to map the results to the interval $(-1,1)$~\cite{Ignatov2021}. The second part consists of a stack of RMFA modules. The third part is the output module, where a convolutional layer followed by a sigmoid activation function is used to generate the output 
\subsection{Black Level Correction}
In the case of an electronic device, the sensor generates electrons even in the absence of light. These electrons will also be collected and readout by the sensor. To address this issue, the sensor incorporates optical black pixels to mitigate the black level and adds a fixed pedestal to the final output. For instance, the IMX586 sensor(see \url{https://www.sony.com/en/SonyInfo/News/Press/201807/18-060E/}) has a black offset of 63. However, the pedestal can introduce color shift, particularly in dim scenes. Instead of directly normalizing the raw data as done in many existing deep learning-based algorithms, we subtract the black level first to address this concern.
\subsection{Three Channel Split}
\begin{figure}[tb]
  \centering
  \includegraphics[width=1\linewidth]{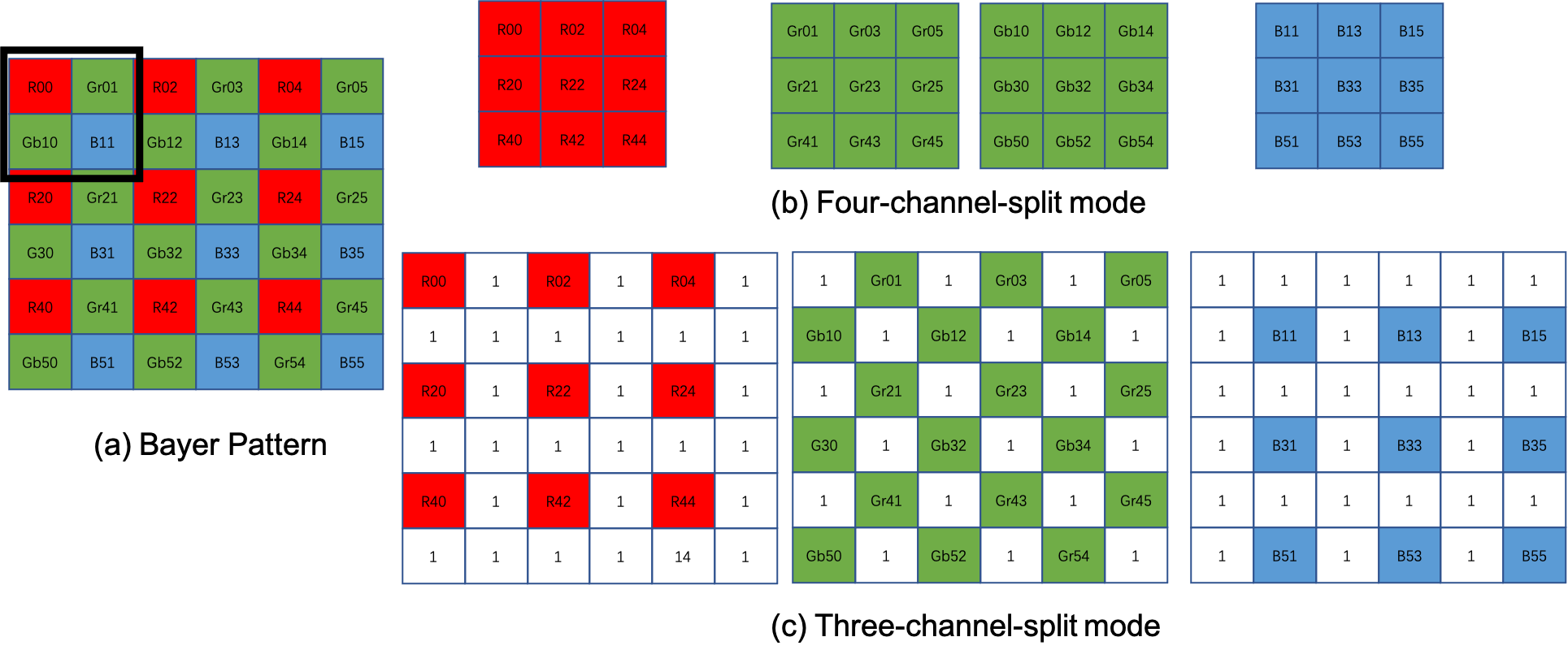}
  \caption{ (a) Bayer CFA; (b) Four-Channel-Split mode;(c) Three-Channel-Split mode
  }
  \label{fig:Bayer}
\end{figure}

In previous works, it is common practice to split each channel (R, Gr, Gb, B) of the Bayer pattern and stack them as the inputs to neural networks, as depicted in \cref{fig:Bayer}(b). However, we argue that this approach may not be optimal for two reasons.

Firstly, each channel of the raw data has different sampling rates. In the Bayer pattern, the green channel has a sampling rate twice that of the red and blue channels, as shown in \cref{fig:Bayer}(a). Additionally, the green channel typically exhibits better sensitivity compared to the red and blue channels, resulting in more textures, especially high-frequency textures, and higher signal-to-noise ratio (SNR). We believe that the green channel is more beneficial for recovering high-frequency textures and should be handled with care.

Secondly, there is a loss of high-frequency information when using the four-channel mode. In this mode, the green channel is further split into Gr and Gb channels. This additional split corresponds to applying downsampling to the green channel, which inevitably leads to the loss of high-frequency information in the raw data. The size of data in the four-channel mode is half that of the original size, as depicted in \cref{fig:Bayer}(b). For example, a $3\times3$ patch of four-channel data corresponds to a $5\times5$ patch in the original-sized data. This artificial change in spatial frequency makes it challenging for the network to accurately extract and reconstruct the lost high-frequency information in the original $5\times5$ patch.

Furthermore, misalignment is another issue to consider. As illustrated by the black box in \cref{fig:Bayer}(a), pixels from the same location in the four channels actually correspond to $2\times2$ neighborhoods in the original raw data. This misalignment is likely to cause blur and negatively impact the reconstruction quality.

To address the aforementioned problems, we have devised a new method, as illustrated in \cref{fig:Bayer}(c).  We split the Bayer raw data into three channels (R, G, B), where each channel retains the size of the raw data. For the unsampled pixels, we fill them with 1s. As a result, the sampling rate of the G channel remains unchanged, preserving the high-frequency texture information as much as possible. We believe this approach will be more beneficial for the network to reconstruct the high-frequency information accurately.
	
It is important to note that this method requires additional computation and memory compared to the traditional four-channel mode. However, we have made efforts to optimize the network architecture to strike a balance between computational cost and memory usage. The goal is to ensure efficient processing while still retaining the benefits of the proposed splitting method.

By adopting this new approach, we aim to address the issues related to sampling rate, high-frequency information loss, and misalignment, ultimately improving the reconstruction quality of the ISP algorithm.

\subsection{Texture Module}
In order to achieve high-quality image reconstruction, it is important to capture both high-frequency and low-frequency textures. To address this, we have designed a texture module that consists of two sub-branches:

\begin{itemize}
	\item [1)] \textbf{High-Frequency Information Extraction Branch}: This sub-branch focuses on extracting high-frequency information from the input data. It utilizes a kernel size of $1\times1$ to capture fine details and subtle variations in the image. By using a smaller kernel size, the network can effectively capture high-frequency textures and preserve the intricate details in the reconstructed image.
	
	\item [2)] \textbf{Low-Frequency Branch}: This sub-branch is responsible for capturing the low-frequency information in the input data. It utilizes a larger kernel size of $3\times3$ to capture broader features and smooth out the image. The larger kernel size allows the network to capture low-frequency textures, such as overall color and tone variations, and ensure the reconstructed image maintains a visually pleasing appearance.
\end{itemize}

By combining these two sub-branches within the texture module, we aim to capture a wide range of textures and enhance the overall image quality during the reconstruction process. The high-frequency branch preserves fine details, while the low-frequency branch captures broader features, resulting in a more comprehensive representation of the image's texture information.

\subsection{Tone Mapping}
To generate 8-bit RGB images from 12-bit raw data, it is crucial for the network to perform tone mapping, which involves adjusting the tone locally and globally to achieve higher contrast while maintaining consistent brightness and color. Existing neural network-based methods typically achieve this implicitly by minimizing certain structure losses. However, these approaches often overlook the uneven exposure property of raw data, which can hinder the learning process.

Raw data is typically captured with auto exposure to obtain suitable brightness. However, the auto exposure strategy may only properly cover certain areas of the image, leaving other regions overexposed or underexposed. This introduces a challenge as the unevenness can vary with ambient illuminance, resulting in diverse tone distributions within the raw data. When learning a tone mapping function, the network is expected to locally brighten underexposed areas, darken overexposed areas, and globally render the tone to adapt to the 8-bit output. Essentially, the network needs to map this diversity to a relatively consistent range, which can be seen as an approximate many-to-one mapping. We argue that this significantly increases the difficulty and complexity of the learning process.

In our approach, we aim to simplify this problem by addressing the uneven exposure directly and transforming the relationship from a many-to-one to a one-to-one mapping. According to Retinex theory~\cite{Land1971}, an image can be decomposed into reflectance and illumination components:
\begin{equation}
  Retinex(I)=Reflectance(r)\cdot Illumination(S)
  \label{eq:retinex}
\end{equation}

The reflectance component represents the intrinsic properties of the image and remains consistent under all lighting conditions, while the illumination component represents the variations in lighting~\cite{Land1971}. We assume that by extracting the consistent reflectance, we can remove the uneven exposure, simplifying the tone mapping function from a many-to-one mapping to a one-to-one mapping.

Based on this analysis, we explicitly designed a tone mapping module. To estimate the illumination component, which relies more on global features, we introduced a Pooling-Convolution structure. This structure downsamples the input data through a pooling layer and then applies a convolutional layer. The initial pooling layer enlarges the receptive field of the subsequent convolutional layer, and this unit can be repeated to extract more accurate global features. The feature maps are then upsampled to their original size. To estimate the reflectance, we directly subtract the estimated illumination from the outputs of the texture module. As described in subsection 3.4, the texture module extracts local textures using small convolutional kernels. By removing the low-frequency information extracted by the global features, we consider the final outputs as the reflectance of the image.

By explicitly considering the uneven exposure property and designing a dedicated tone mapping module, we aim to improve the learning process and enhance the network's ability to perform effective tone mapping while generating 8-bit RGB images from 12-bit raw data
\subsection{RMFA  block}
The Residual Multi-Frequency Attention (RMFA) module serves as the fundamental building block of our model. As depicted in \cref{fig:networks}(b), the outputs of the texture module and tone mapping module are first concatenated together. Subsequently, a convolutional layer is employed to map the number of feature maps to the original depth width. ReLU activation function is applied after each convolutional operation. Channel attention and spatial attention modules are then sequentially added. Finally, a skip connection is introduced to prevent performance degradation. Being a versatile building block, the RMFA module can be seamlessly integrated into various architectures, enhancing the flexibility and adaptability of our model.
\section{Experiment and Results}
\label{sec:experiment}
\subsection{Dataset}

The Mobile AI Workshop~\cite{Ignatov2023a, Ignatov2021} provided a dataset consisting of 24,000 pairs of training RAW-RGB images for algorithm development. In this paper, we utilized this dataset for both training and evaluating our proposed model. Originally, this dataset was part of the Fujifilm UltraISP dataset~\cite{Ignatov2022}, which involved capturing high-quality images using the Fujifilm GFX100 medium format 102 MP camera and acquiring raw data from the Sony IMX586 Quad Bayer mobile camera sensor. As the collected RAW-RGB image pairs were not perfectly aligned, a state-of-the-art deep learning-based dense matching algorithm was employed to align the images. From the aligned images, $256\times256$ pixel patches were extracted for further processing. It is important to note that all alignment operations were conducted solely on the Fujifilm RGB images, while the RAW data from the Sony sensor remained unmodified, preserving the original values as read from the sensor.

To create train and test datasets, we randomly divided the dataset with a ratio of 9:1, ensuring a stable evaluation. We repeated this experiment 10 times and the reported results are the average output of each repetition, providing robustness and reliability in our findings. 

\subsection{Training Details}
The loss function employed for training the RMFA-NET model is a linear combination of several components, including the L1 loss, perception loss based on four VGG16 layers~\cite{Johnson2016}, structural similarity index measure (SSIM) loss~\cite{Wang2004}, and color loss~\cite{Ignatov2017}. The L1 loss is minimized with a weight parameter $\theta$ set to 1.0, while the weights for the other components (perception, SSIM, and color) are adjusted accordingly. The overall loss function is defined as follows:
\begin{equation}
  Loss = \theta L_{l1}+\eta L_{vgg}+\lambda L_{ssim}+\gamma L_{color}
  \label{eq:loss}
\end{equation}

The model was implemented using PyTorch and trained on a single NVIDIA Tesla A100 GPU. The batch size varied between 8 and 64, depending on the scale of the model. During training, image patches were randomly extracted from the RAW images, with dimensions of $3\times256\times256$ using three-channel packing (RGB). Corresponding patches were also extracted from the sRGB images with the same dimensions.

The model parameters were optimized for 1000 epochs using the Adam algorithm~\cite{Kingma2014} with an initial learning rate of 1e-3. A cosine annealing schedule~\cite{Loshchilov2016} was employed, aggressively decreasing the learning rate to 1e-7 every 100 epochs.

\subsection{Ablation Study}
We compared our method with the top 5 winners of the Mobile AI Workshop@CVPR 2022~\cite{Ignatov2023a} in terms of visual results. The workshop allowed participants to develop larger and more powerful models for the task at hand. To evaluate the image quality, we used the PSNR and SSIM metrics, which were also employed by the workshop. Additionally, we provide the parameters of each model for reference
\begin{table}[htb]
  \caption{Average PSNR (Peak Signal-to-Noise Ratio) and SSIM (Structural Similarity Index Measure) results on the test images.
  }
  \label{tab:ablation_study}
  \centering
  \begin{tabular}{@{}llll@{}}
    \toprule
    Channel Mode & PSNR & SSIM &   Parameter (MB)\\
    \midrule

    {\bf Ours RMFA-Net} & {\bf 25.1} & {\bf 0.889} & 0.19 \\
			HITZST01 & 24.09 & 0.8667 & 1.2 \\
			ENERZAi & 24.08 & 0.8778 & 4.5 \\
			CASIA 1st  & 24.09 & 0.884 & 205 \\
			Multimedia  & 23.96 & 0.8543 & 0.029 \\
			HITZST01  & 23.89 & 0.8666 & 0.060 \\
  \bottomrule
  \end{tabular}
\end{table}

\begin{figure*}[tb]
	\centering
	\setlength{\tabcolsep}{1pt}
	\resizebox{\linewidth}{!}
	{
		\begin{tabular}{cccccc}
			\includegraphics[width=0.33\linewidth]{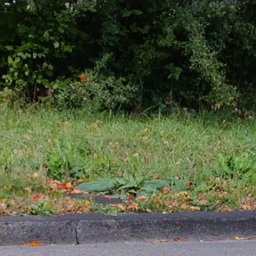}&
			\includegraphics[width=0.33\linewidth]{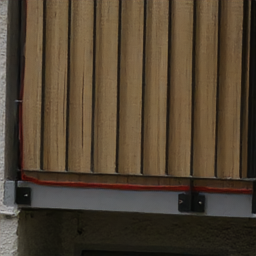}&
			\includegraphics[width=0.33\linewidth]{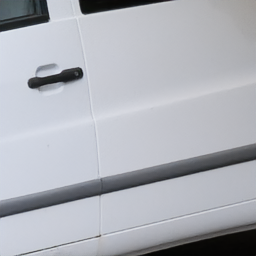}&
			\includegraphics[width=0.33\linewidth]{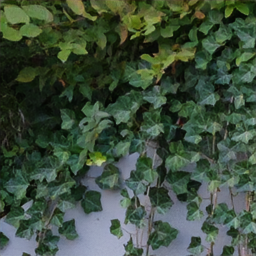}&
			\includegraphics[width=0.33\linewidth]{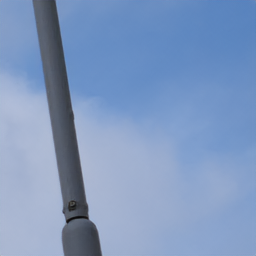}&
			\includegraphics[width=0.33\linewidth]{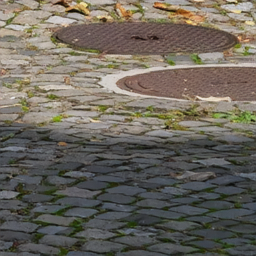} \\
			%%%%%%%%%%%%%%%%%%%%%%%%%%%%%%%%%%%%%%%%%%%%%%%%%%%%%%%%%%%%%%%%%%%%%%%%%%%%%%
			\includegraphics[width=0.33\linewidth]{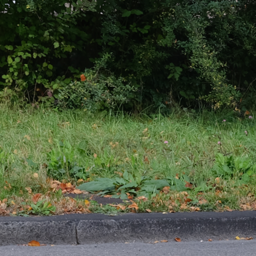}&
			\includegraphics[width=0.33\linewidth]{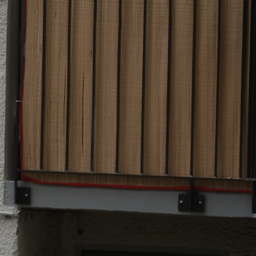}&
			\includegraphics[width=0.33\linewidth]{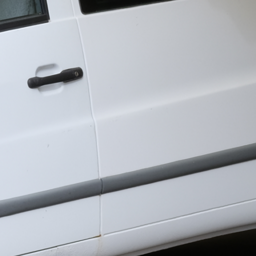}& 
			\includegraphics[width=0.33\linewidth]{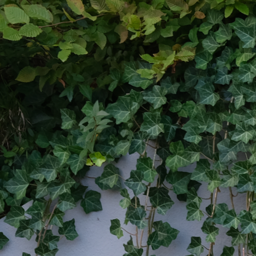}&
			\includegraphics[width=0.33\linewidth]{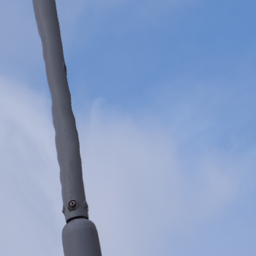}&
			\includegraphics[width=0.33\linewidth]{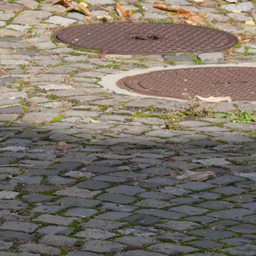} \\
		\end{tabular}
	}
	\caption{Outputs of RMFA-Net (top) and target photos captured with the Fujifilm GFX100 (bottom).}
	\label{fig:ablation_study}
\end{figure*}
As shown in \cref{tab:ablation_study}, our method achieved the highest results in terms of PSNR and SSIM. \cref{fig:ablation_study} displays the outputs of RMFA-Net alongside the corresponding target photos. It is evident that the network's outputs are accurate and closely resemble the ground truth. Both global and local levels of brightness are reconstructed with precision. The estimation of hue and saturation of colors is also commendable, exhibiting no noticeable color cast. The white balance is successfully achieved. Considering that the dataset was collected during daytime with sunlight as the only illuminant, it should be noted that when designing a neural ISP algorithm with a more powerful and robust white balance function, it is important to consider a wider range of illuminant types and color temperatures (CCT).

The recovery of resolution and texture is consistent with expectations, particularly in high-frequency areas. Moreover, no prominent artifacts, such as false color or zipper effects, are observed in these regions. The overall contrast and sharpness of the output images are also commendable, indicating successful learning of the tone mapping function.
\subsection{The Effectiveness of Three Channel Split Mode}

To demonstrate the effectiveness of the channel split approach proposed in this paper, we designed a four-channel version. The raw data is preprocessed following the method used in PyNet~\cite{Ignatov2020}, and pixel-shuffle operations are employed to upsample the network outputs to the original size.

\begin{table}[htb]
  \caption{The quantitative results of RMFA-NET and RMFA-Net-four-channel-split, along with the winner of Mobile AI Workshop@CVPR 2022~\cite{Ignatov2023a}, are presented for reference.
  }
  \label{tab:channel_split}
  \centering
  \begin{tabular}{@{}llll@{}}
    \toprule
    Channel Mode & PSNR & SSIM &   Parameter (MB)\\
    \midrule
    {\bf RMFA-Net} & {\bf 25.1} & {\bf 0.889} & 0.19 \\
    {\bf RMFA-Net-Four} & {\bf 24.7} & {\bf 0.881} & 0.196 \\
    HITZST01 & 24.09 & 0.8667 & 1.2 \\
  \bottomrule
  \end{tabular}
\end{table}

\begin{figure*}[tb]
	\centering
	\setlength{\tabcolsep}{1pt}
	\resizebox{\linewidth}{!}
	{
		\begin{tabular}{ccc}
			\scriptsize{Four-channel mode}\normalsize & \scriptsize{Three-channel mode}\normalsize & \scriptsize{Fujifilm GFX 100 Target Photo}\normalsize\\
			\includegraphics[width=0.33\linewidth]{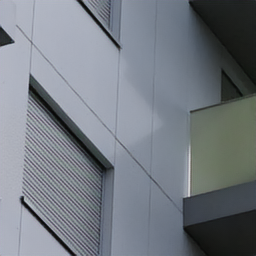}&
			\includegraphics[width=0.33\linewidth]{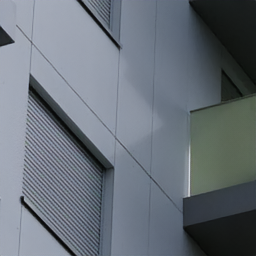}&
			\includegraphics[width=0.33\linewidth]{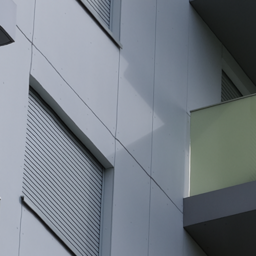} \\
			\includegraphics[width=0.33\linewidth]{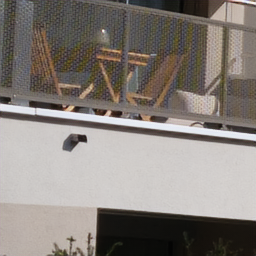}&
			\includegraphics[width=0.33\linewidth]{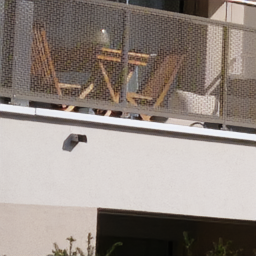}&
			\includegraphics[width=0.33\linewidth]{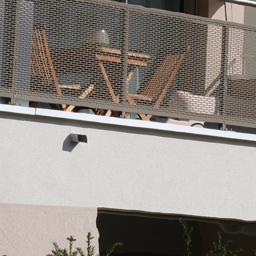} \\
			
		\end{tabular}
	}
	\caption{Outputs of four-channel mode (left), three-channel mode (middle), and target photos captured with the Fujifilm GFX100 (right).}
	\label{fig:channel_split}
\end{figure*}
From \cref{tab:channel_split}, it can be observed that the channel split method outperforms the four-channel approach by approximately 0.4dB. \cref{fig:channel_split} illustrates the visual results of each model alongside their corresponding ground truth images. In the output images of the four-channel model, noticeable artifacts such as moire patterns, blurring, zipper effects, and false colors are evident. Conversely, these artifacts are barely noticeable in the outputs of RMFA-Net, highlighting the effectiveness of the three-channel-split mode.

In the four-channel-split mode, the input data size is half of the raw data. To reconstruct the original size, upsampling operations are necessary. Pixel shuffle and interpolation are commonly used for this purpose. However, it is important to note that there is a natural limitation on the sampling rate in order to accurately recover the original signal. According to the Nyquist Sampling Theorem~\cite{Landau1967}, a bandlimited continuous-time signal can be perfectly reconstructed from its samples if the waveform is sampled at a rate that is at least twice as fast as its highest frequency component. In other words, the highest frequency component that can be accurately 
represented is limited to:
\begin{equation}
  f_{max}<\frac{1}{2}f_{s}
  \label{eq:loss}
\end{equation}
Frequencies higher than this limit cannot be accurately recovered and will appear as low-frequency aliasing. This is why moire patterns or zipper effects are more likely to occur in the four-channel mode, especially in areas with high frequencies. To remove these effects, more complex network architectures or post-processing techniques would be required. This explains the observed PSNR drop in the four-channel mode models.

\subsection{The Effectiveness of Tone Mapping Module}

\begin{figure*}[tb]
	\centering
	\setlength{\tabcolsep}{1pt}
	\resizebox{\linewidth}{!}
	{
		\begin{tabular}{ccccc}
			\includegraphics[width=0.33\linewidth]{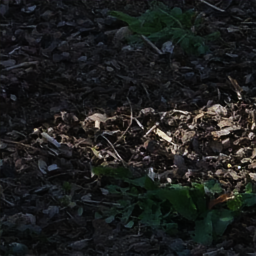}&
			\includegraphics[width=0.33\linewidth]{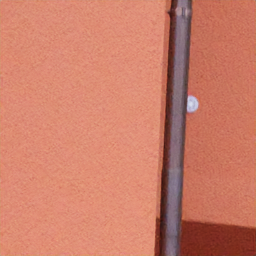}&
			\includegraphics[width=0.33\linewidth]{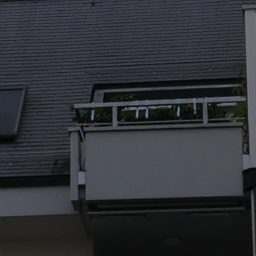}&
			\includegraphics[width=0.33\linewidth]{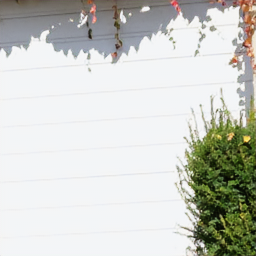}&
			\includegraphics[width=0.33\linewidth]{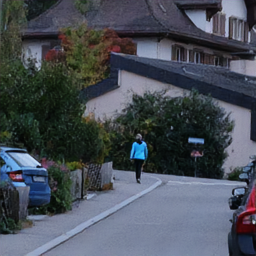} \\
			%%%%%%%%%%%%%%%%%%%%%%%%%%%%%%%%%%%%%%%%%%%%%%%%%%%%%%%%%%%%%%%%%%%%%%%%%%%%%%
			\includegraphics[width=0.33\linewidth]{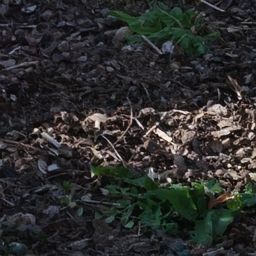}&
			\includegraphics[width=0.33\linewidth]{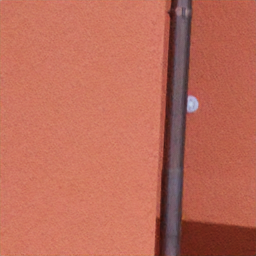}&
			\includegraphics[width=0.33\linewidth]{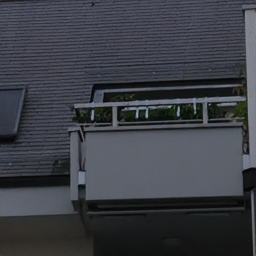}&
			\includegraphics[width=0.33\linewidth]{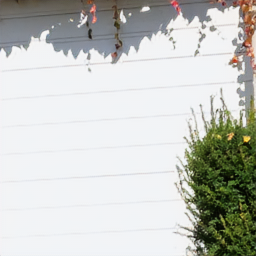}&
			\includegraphics[width=0.33\linewidth]{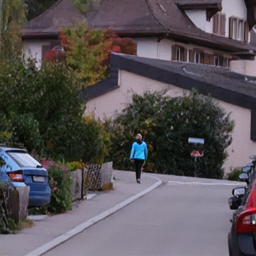} \\
			%%%%%%%%%%%%%%%%%%%%%%%%%%%%%%%%%%%%%%%%%%%%%%%%%%%%%%%%%%%%%%%%%%%%%%%%%%%%%%
			\includegraphics[width=0.33\linewidth]{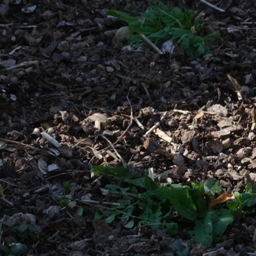}&
			\includegraphics[width=0.33\linewidth]{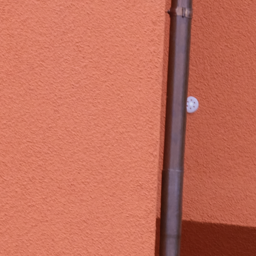}&
			\includegraphics[width=0.33\linewidth]{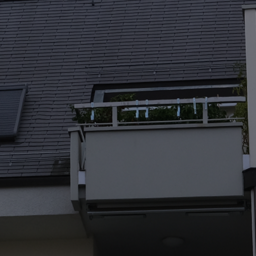}& 
			\includegraphics[width=0.33\linewidth]{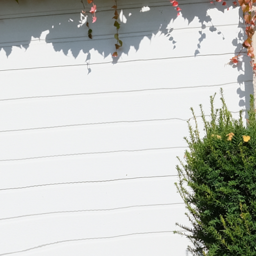}&
			\includegraphics[width=0.33\linewidth]{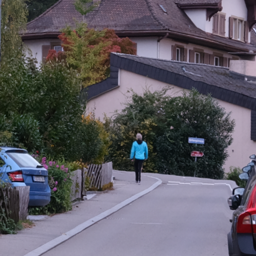} \\
		\end{tabular}
	}
	\caption{Outputs of RMFA-without-tonemapping (top row), RMFA-NET (middle row), and Fujifilm GFX100 target photos (bottom row).}
	\label{fig:tm}
\end{figure*}

To demonstrate the effectiveness of the tone mapping module, we conducted an experiment where we removed it from RMFA-Net. The results are reported in  \cref{tab:tm}, revealing a decrease in PSNR of over 0.2dB.  \cref{fig:tm} visually presents the impact of the tone mapping module, showing that it achieves a high level of consistency with the target photos. The outputs of RMFA-Net with the tone mapping module exhibit better contrast and sharpness. The brightness is accurately reconstructed at a similar level to the target photos.

\begin{table}[htb]
  \caption{The results of RMFA-NET and RMFA-Net-without-tonemapping. Winner of Mobile AI Workshop@CVPR 2022~\cite{Ignatov2023a} is provided for the reference.
  }
  \label{tab:tm}
  \centering
  \begin{tabular}{@{}llll@{}}
    \toprule
    Channel Mode & PSNR & SSIM &   Parameter (MB)\\
    \midrule
    {\bf RMFA-Net} & {\bf 25.1} & {\bf 0.889} & 0.19 \\
    {\bf RMFA-Net-without-tonemapping} & {\bf 24.87} & {\bf 0.886} & 0.196 \\
    HITZST01 & 24.09 & 0.8667 & 1.2 \\
  \bottomrule
  \end{tabular}
\end{table}

By comparing the first and third columns of \cref{fig:tm}, it is evident that without the tone mapping module, the network fails to accurately recover information in dim scenes. Additionally, in the absence of the module, the highlight areas become prone to overexposure, leading to a loss of details as depicted in the fourth column. With the inclusion of the tone mapping module and accurate brightness adjustment, the saturation is reconstructed more correctly, as shown in the second and fifth columns. The absence of the tone mapping module results in an obvious color cast in the network outputs, indicating that the explicitly designed tone mapping module is more effective in achieving accurate and consistent color representation.

\subsection{Adjusting the Model Complexity}

RMFA-NET offers the flexibility to adapt its computational complexity by adjusting the number of building blocks. This allows for the design of networks with varying sizes based on the desired runtime, task performance, and computational budget. In this section, we present five RMFA-Net models with different depths and widths. Specifically, for the tiny model, we stack two RMFA blocks with a channel width of 16. For the medium model, we use 8 RMFA blocks with a channel width of 16. The large model consists of 20 RMFA blocks with a channel width of 16. Additionally, we extend the channel width of the large model to 32 and 64. The quantitative results are presented in \cref{tab:mc}, indicating that even the tiny model achieves a commendable score. This demonstrates the flexibility of the model's architecture, allowing for the adjustment of complexity to accommodate devices with varying computational power.

\begin{table}[tb]
  \caption{Quantitative results of several RMFA-NET models with different depth and width. Winner of Mobile AI Workshop@CVPR 2022~\cite{Ignatov2023a} is provided for the reference.
  }
  \label{tab:mc}
  \centering
  \begin{tabular}{@{}llll@{}}
    \toprule
    Channel Mode & PSNR & SSIM &   Parameter (MB)\\
    \midrule
    {\bf Tiny-RMFA-Net-W16} & {\bf 24.549} & {\bf 0.88} & 0.022 \\
			{\bf Medium-RMFA-Net-W16} & {\bf 24.86} & {\bf 0.8819} & 0.0798 \\
			{\bf Large-RMFA-Net-W16}  & {\bf 25.1} & {\bf 0.889} & 0.19 \\
			{\bf Large-RMFA-Net-W32}  & {\bf 25.315} & {\bf 0.8911} & 0.7726 \\
			{\bf Large-RMFA-Net-W64}  & {\bf 25.22 }& {\bf 0.8906 }& 3.16 \\
    HITZST01 & 24.09 & 0.8667 & 1.2 \\
  \bottomrule
  \end{tabular}
\end{table}
\subsection{Limitations}
While RMFA-Net shows significant improvements in RAW to RGB image reconstruction, there are still several avenues for future exploration and improvement. We identify the following limitations and challenges:

\textbf{Generalization to diverse RAW inputs:} Our model is trained and evaluated on a specific dataset with limited variations in illuminants and color temperatures. Future work should focus on enhancing the generalization ability of RMFA-Net to handle a wider range of RAW inputs captured in various lighting conditions and scenes.
	
	\textbf{Computational efficiency:} While RMFA-Net achieves impressive results in terms of image quality, it is important to note that it comes with a higher computational cost. This increased computational cost is primarily introduced by the three-channel split mode, which keeps the input data at its original size. Balancing the computational cost and the desired level of image quality is an important consideration when deploying RMFA-Net in practical applications. Moreover, exploring methods to optimize the model's architecture or develop lightweight versions without compromising performance would be beneficial, especially for real-time applications on resource-constrained devices.
	
	\textbf{Robustness to noise and artifacts:} RAW images often contain noise and artifacts due to sensor limitations or imperfect alignment. Enhancing the robustness of RMFA-Net to handle such challenges and produce clean and artifact-free outputs is an important area for future research.

\section{Conclusion}
\label{sec:conclusion}
In this paper, we proposed RMFA-Net, a novel deep learning model for real RAW to RGB image reconstruction. Through extensive experiments and evaluations, we demonstrated the effectiveness of our approach in terms of quantitative metrics such as PSNR and SSIM, as well as visual quality compared to state-of-the-art methods. RMFA-Net excels in reconstructing accurate brightness, color, texture, and overall image details, providing promising results for real-world RAW image processing.

While RMFA-Net shows significant advancements, there are still several areas for future exploration and enhancement. One key aspect is optimizing computational efficiency, which is crucial for practical deployment. Additionally, improving the robustness of the model would be beneficial. Considering the diversity of sensors, it is important to enhance its generalization in future research.

\bibliographystyle{splncs04}
\bibliography{rmfa_net}

\begin{thebibliography}{10}
\providecommand{\url}[1]{\texttt{#1}}
\providecommand{\urlprefix}{URL }
\providecommand{\doi}[1]{https://doi.org/#1}

\bibitem{Bayer1976}
Bayer, B.E.: Color imaging array (1976)

\bibitem{Cai2016}
Cai, B., Xu, X., Jia, K., Qing, C., Tao, D.: Dehazenet: An end-to-end system for single image haze removal. IEEE Transactions on Image Processing  \textbf{25}(11),  5187--5198 (2016)

\bibitem{Cai2018}
Cai, J., Gu, S., Zhang, L.: Learning a deep single image contrast enhancer from multi-exposure images. IEEE Transactions on Image Processing  \textbf{27}(4),  2049--2062 (2018)

\bibitem{Chen2018}
Chen, C., Chen, Q., Xu, J., Koltun, V.: Learning to see in the dark. In: 2018 IEEE/CVF Conference on Computer Vision and Pattern Recognition (2018)

\bibitem{Cho2021}
Cho, S.J., Ji, S.W., Hong, J.P., Jung, S.W., Ko, S.J.: Rethinking coarse-to-fine approach in single image deblurring (2021)

\bibitem{Dong2015}
Dong, C., Loy, C.C., He, K., Tang, X.: Image super-resolution using deep convolutional networks. IEEE transactions on pattern analysis and machine intelligence  \textbf{38}(2),  295--307 (2015)

\bibitem{Guo2021}
Guo, Q., Sun, J., Juefei-Xu, F., Ma, L., Xie, X., Feng, W., Liu, Y., Zhao, J.: Efficientderain: Learning pixel-wise dilation filtering for high-efficiency single-image deraining. In: Proceedings of the AAAI Conference on Artificial Intelligence. vol.~35, pp. 1487--1495 (2021)

\bibitem{Hu2019}
Hu, X., Naiel, M.A., Wong, A., Lamm, M., Fieguth, P.: Runet: A robust unet architecture for image super-resolution. In: IEEE Conference on Computer Vision and Pattern Recognition Workshops (2019)

\bibitem{Ignatov2021}
Ignatov, A., Chiang, C.M., Kuo, H.K., Sycheva, A., Timofte, R.: Learned smartphone isp on mobile npus with deep learning, mobile ai 2021 challenge: Report. In: Proceedings of the IEEE/CVF Conference on Computer Vision and Pattern Recognition. pp. 2503--2514 (2021)

\bibitem{Ignatov2017}
Ignatov, A., Kobyshev, N., Timofte, R., Vanhoey, K., Van~Gool, L.: Dslr-quality photos on mobile devices with deep convolutional networks. In: Proceedings of the IEEE international conference on computer vision. pp. 3277--3285 (2017)

\bibitem{Ignatov2022}
Ignatov, A., Malivenko, G., Timofte, R., Tseng, Y., Xu, Y.S., Yu, P.H., Chiang, C.M., Kuo, H.K., Chen, M.H., Cheng, C.M., et~al.: Pynet-v2 mobile: Efficient on-device photo processing with neural networks. In: 2022 26th International Conference on Pattern Recognition (ICPR). pp. 677--684. IEEE (2022)

\bibitem{Ignatov2023}
Ignatov, A., Sycheva, A., Timofte, R., Tseng, Y., Xu, Y.S., Yu, P.H., Chiang, C.M., Kuo, H.K., Chen, M.H., Cheng, C.M., et~al.: Microisp: Processing 32mp photos on mobile devices with deep learning. In: Computer Vision--ECCV 2022 Workshops: Tel Aviv, Israel, October 23--27, 2022, Proceedings, Part II. pp. 729--746. Springer (2023)

\bibitem{Ignatov2023a}
Ignatov, A., Timofte, R., Liu, S., Feng, C., Bai, F., Wang, X., Lei, L., Yi, Z., Xiang, Y., Liu, Z., et~al.: Learned smartphone isp on mobile gpus with deep learning, mobile ai \& aim 2022 challenge: report. In: Computer Vision--ECCV 2022 Workshops: Tel Aviv, Israel, October 23--27, 2022, Proceedings, Part III. pp. 44--70. Springer (2023)

\bibitem{Ignatov2020}
Ignatov, A., Van~Gool, L., Timofte, R.: Replacing mobile camera isp with a single deep learning model. In: Proceedings of the IEEE/CVF Conference on Computer Vision and Pattern Recognition Workshops. pp. 536--537 (2020)

\bibitem{Isola2016}
Isola, P., Zhu, J.Y., Zhou, T., Efros, A.A.: Image-to-image translation with conditional adversarial networks. IEEE  (2016)

\bibitem{Jia2021}
Jia, F., Wong, W.H., Zeng, T.: Ddunet: Dense dense u-net with applications in image denoising. In: International Conference on Computer Vision (2021)

\bibitem{Johnson2016}
Johnson, J., Alahi, A., Fei-Fei, L.: Perceptual losses for real-time style transfer and super-resolution. In: Computer Vision--ECCV 2016: 14th European Conference, Amsterdam, The Netherlands, October 11-14, 2016, Proceedings, Part II 14. pp. 694--711. Springer (2016)

\bibitem{Kim2020}
Kim, B.H., Song, J., Ye, J.C., Baek, J.: Pynet-ca: enhanced pynet with channel attention for end-to-end mobile image signal processing. In: Computer Vision--ECCV 2020 Workshops: Glasgow, UK, August 23--28, 2020, Proceedings, Part III 16. pp. 202--212. Springer (2020)

\bibitem{Kim2016}
Kim, J., Lee, J.K., Lee, K.M.: Accurate image super-resolution using very deep convolutional networks. In: Proceedings of the IEEE conference on computer vision and pattern recognition. pp. 1646--1654 (2016)

\bibitem{Kingma2014}
Kingma, D.P., Ba, J.: Adam: A method for stochastic optimization. arXiv preprint arXiv:1412.6980  (2014)

\bibitem{Kong2022}
Kong, F., Li, M., Liu, S., Liu, D., He, J., Bai, Y., Chen, F., Fu, L.: Residual local feature network for efficient super-resolution  (2022)

\bibitem{Kupyn2018}
Kupyn, O., Budzan, V., Mykhailych, M., Mishkin, D., Matas, J.: Deblurgan: Blind motion deblurring using conditional adversarial networks. In: Proceedings of the IEEE conference on computer vision and pattern recognition. pp. 8183--8192 (2018)

\bibitem{Land1971}
Land, E.H., McCann, J.J.: Lightness and retinex theory. Josa  \textbf{61}(1),  1--11 (1971)

\bibitem{Landau1967}
Landau, H.: Sampling, data transmission, and the nyquist rate. Proceedings of the IEEE  \textbf{55}(10),  1701--1706 (1967)

\bibitem{Liu2020}
Liu, L., Jia, X., Liu, J., Tian, Q.: Joint demosaicing and denoising with self guidance. In: Proceedings of the IEEE/CVF Conference on Computer Vision and Pattern Recognition. pp. 2240--2249 (2020)

\bibitem{Loshchilov2016}
Loshchilov, I., Hutter, F.: Sgdr: Stochastic gradient descent with warm restarts. arXiv preprint arXiv:1608.03983  (2016)

\bibitem{Lukac2005}
Lukac, R., Plataniotis, K.N.: Color filter arrays: design and performance analysis. Consumer Electronics IEEE Transactions on  \textbf{51}(4),  1260--1267 (2005)

\bibitem{Pan2017}
Pan, J., Canton, C., Mcguinness, K., O'Connor, N.E., Giro-I-Nieto, X.: Salgan: Visual saliency prediction with generative adversarial networks  (2017)

\bibitem{Ronneberger2015}
Ronneberger, O., Fischer, P., Brox, T.: U-net: Convolutional networks for biomedical image segmentation. In: Medical Image Computing and Computer-Assisted Intervention--MICCAI 2015: 18th International Conference, Munich, Germany, October 5-9, 2015, Proceedings, Part III 18. pp. 234--241. Springer (2015)

\bibitem{Schwartz2018}
Schwartz, E., Giryes, R., Bronstein, A.M.: Deepisp: Toward learning an end-to-end image processing pipeline. IEEE Transactions on Image Processing  \textbf{28}(2),  912--923 (2018)

\bibitem{Shi2016a}
Shi, W., Caballero, J., Huszár, F., Totz, J., Wang, Z.: Real-time single image and video super-resolution using an efficient sub-pixel convolutional neural network. In: 2016 IEEE Conference on Computer Vision and Pattern Recognition (CVPR) (2016)

\bibitem{Uchida2018}
Uchida, K., Tanaka, M., Okutomi, M.: Non-blind image restoration based on convolutional neural network. In: 2018 IEEE 7th Global Conference on Consumer Electronics (GCCE) (2018)

\bibitem{Uhm2019}
Uhm, K.H., Kim, S.W., Ji, S.W., Cho, S.J., Hong, J.P., Ko, S.J.: W-net: Two-stage u-net with misaligned data for raw-to-rgb mapping. In: 2019 IEEE/CVF International Conference on Computer Vision Workshop (ICCVW). pp. 3636--3642 (2019). \doi{10.1109/ICCVW.2019.00448}

\bibitem{Wang2018}
Wang, C., Xu, C., Wanga, C., Tao, D.: Perceptual adversarial networks for image-to-image transformation. IEEE Transactions on Image Processing pp. 4066--4079 (2018)

\bibitem{Wang2021}
Wang, L., Yoon, K.J.: Deep learning for hdr imaging: State-of-the-art and future trends. IEEE transactions on pattern analysis and machine intelligence  \textbf{44}(12),  8874--8895 (2021)

\bibitem{Wang2020}
Wang, Z., Chen, J., Hoi, S.C.: Deep learning for image super-resolution: A survey. IEEE transactions on pattern analysis and machine intelligence  \textbf{43}(10),  3365--3387 (2020)

\bibitem{Wang2004}
Wang, Z., Bovik, A.C., Sheikh, H.R., Simoncelli, E.P.: Image quality assessment: from error visibility to structural similarity. IEEE transactions on image processing  \textbf{13}(4),  600--612 (2004)

\bibitem{Zhang2017}
Zhang, K., Zuo, W., Chen, Y., Meng, D., Zhang, L.: Beyond a gaussian denoiser: Residual learning of deep cnn for image denoising. IEEE transactions on image processing  \textbf{26}(7),  3142--3155 (2017)

\bibitem{Zhang2018}
Zhang, Y., Li, K., Li, K., Wang, L., Zhong, B., Fu, Y.: Image super-resolution using very deep residual channel attention networks. In: Proceedings of the European conference on computer vision (ECCV). pp. 286--301 (2018)

\end{thebibliography}
\end{document}